\documentclass[prl,twocolumn,showpacs,floatfix,amsfonts]{revtex4}
\usepackage{graphicx,graphics,color,epsfig,exscale}
\usepackage{bm}
\usepackage{amsmath}
\usepackage{amssymb}

\def\sgn{\text{sgn}}

\def\be{\begin{equation}} 
\def\ee{\end{equation}} 
\def\bea{\begin{eqnarray}} 
\def\eea{\end{eqnarray}}
\begin{document}
\preprint{}
\title{Berry Phase and the Breakdown of the Quantum to Classical
Mapping for the Quantum Critical Point of the Bose-Fermi Kondo model}
\author{Stefan Kirchner and Qimiao Si}
\affiliation
{Department of Physics \& Astronomy, Rice University, Houston, 
TX 77005, USA}

\begin{abstract}
The phase diagram of the Bose-Fermi Kondo model contains an SU(2)-invariant 
Kondo-screened phase separated by a continuous quantum phase transition 
from a Kondo-destroyed local moment phase. We analyze  the 
effect of the Berry phase term of the spin path integral on the
quantum critical properties of this quantum impurity model.
For a range of the power-law exponent characterizing the spectral density 
of the dissipative bosonic bath, 
neglecting the influence of the Berry phase term makes the fixed point 
Gaussian.
For the same range of the spectral density exponent,
incorporating the Berry phase term leads instead 
to an interacting fixed point,
for which a quantum to classical mapping breaks down.
Some general implications of our results are discussed.
\end{abstract}
\pacs{71.10.Hf, 05.70.Jk, 75.20.Hr, 71.27.+a}
\maketitle
Quantum criticality has become a new paradigm in the study of the overall
phase diagram of strongly correlated electron systems.
Universal properties of a quantum critical point (QCP) are traditionally
described in terms of a mapping to
the classical critical fluctuations of an order parameter
in elevated dimensions~\cite{Sachdev}.
The quantum to classical mapping is based on the notion
that slow fluctuations of the order parameter are the only
critical degrees of freedom.
This notion has been challenged in a number of contexts.
In the heavy fermion metals,
an anti-ferromagnetic QCP can accommodate new quantum modes,
which are characterized by a critical destruction of the 
Kondo effect~\cite{Si.01,Coleman.01,Loehneysen.07,Gegenwart.08}.

In addition to lattice Kondo systems, the critical Kondo destruction
has also been studied in the Bose-Fermi Kondo model 
(BFKM)~\cite{Zhu.02,Zarand.02}.
The first indication for the violation of the quantum to classical
mapping came from a study in a dynamical large-$N$ limit
of the spin-isotropic BFKM~\cite{Zhu.04}.
For the spectral density of the bosonic bath of the 
form $|\omega|^{1-\epsilon}\sgn \omega$, 
the large-$N$ limit yields an interacting 
fixed point not only for $0 < \epsilon < 1/2$ but
also for $1/2 \le \epsilon < 1$.
Related conclusions were drawn based on the numerical renormalization 
group (NRG) studies
of the related quantum impurity models with
Ising anisotropy~\cite{Vojta.05,Glossop.05}. 
Very recently, the results of the Ising anisotropic models have been 
the subject
of renewed interest, in light of the contrasting behavior between 
the NRG results
and those from Monte Carlo simulations of a classical Ising 
chain~\cite{Kirchner.08b,Winter.08,Guidon.08}.

Given these recent developments, it is timely to address the
issue of the quantum to classical mapping in the spin-isotropic BFKM.
For this purpose, we consider the model in terms of a coherent-state
spin path
integral representation, which highlights the role of the Berry phase term.
By separately considering the cases in the presence/absence 
of the Berry phase term, we establish that the breakdown 
of the quantum to classical mapping 
originates from the interference effect 
of the Berry phase term.

{\it The spin-isotropic Bose-Fermi Kondo model:~~}
The model is specified by the Hamiltonian,
\begin{eqnarray}
{\cal H}_{\text{\small bfkm}} &=& j_K ~{\bf S}
\cdot {\bf s}_c + \sum_{p\sigma}
E_{p}~c_{p\sigma}^{\dagger}~ c_{p\sigma}
\nonumber\\
&+& \; 
g_0
\sum_{p} {\bm S} \cdot \left( {\bm \phi}_{p} + {\bm \phi}_{-p}^{\;\dagger}
\right) + \sum_{p} w_{p}\,{\bm \phi}_{p}^{\;\dagger} {\bm \phi}_{p}\; .
\label{EQ:H-imp}
\end{eqnarray}
Here ${\bf S}$ is a spin-$1/2$ local moment, 
$j_K$ and $g_0$ 
are the Kondo coupling 
and the coupling constant to the bosonic bath respectively,
$c_{p\sigma}^{\dagger}$ describes a fermionic bath
with a constant density
of states, $\sum_{p} \delta (\omega - E_{p}) = N_0$,
and ${\bm \phi}_{p}^{\;\dagger}$ is the  bosonic bath
with the spectral density:
\begin{eqnarray}
\mbox{Im}\chi_0^{-1}(\omega)\,&\equiv&\,\sum_p [\delta(\omega-w_p)- \delta(\omega+w_p)]\nonumber \\ 
&\,\sim&\,
|\omega|^{1-\epsilon} \sgn(\omega) \Theta(\omega_c-|\omega|).
\label{EQ:sub-Ohmic}
\end{eqnarray}
The 
partition function 
of this model is
\begin{eqnarray}
{\mathcal Z}\,&=&\, \int {\mathcal
  D}[\bar{c}_{\sigma}^{},c_{\sigma}^{},{\bm \bar{\phi}},{\bm
  {\phi}}^{},\vec{n}]\, \delta(|\vec{n}|^2-1)\nonumber \\ 
&&~~~~~~~\exp[{-S(\bar{c}_{\sigma}^{ },c_{\sigma}^{},{\bm \bar{\phi}},{\bm
  {\phi}}^{ },\vec{n})}],
\end{eqnarray}
where $c_{\sigma}$ is a Grassmann variable for the fermionic coherent
state, while $ \phi$ and $\vec{n}$ are c-numbers for the
bosonic and spin coherent states respectively.
The action is given by
\begin{eqnarray}
S &=&\, i s [\omega(\vec{n})]+\int_{0}^{\beta} d\tau\,
\sum_p (\Sigma_\sigma
\bar{c}_{p\sigma}^{ } \partial_{\tau} c_{p\sigma}^{} +
{\bm \bar{\phi}_p^{ }}\partial_{\tau}{\bm \phi}_p)\nonumber \\
 &+&\, \int_{0}^{\beta}
d\tau\,
H_{\mbox{\tiny bfkm}}(\bar{c}_{}^{ },c_{}^{},{\bm \bar{\phi}},{\bm
  {\phi}}^{ },s\vec{n}).
\label{eq:BFKM-action}
\end{eqnarray}

{\it Berry phase of the coherent-state spin path integral:~~}
Eq.~(\ref{eq:BFKM-action}) is written in terms of the 
spin (or SU(2)) coherent states~\cite{Perelomov,Inomata}.
Just like bosonic coherent states, spin coherent states are generated by 
a unitary operator
\begin{equation}
|\xi>=T(\xi)|0>;~T(\xi)=e^{[\xi J^\dagger - \bar{\xi} J^-]}
\end{equation}
acting on a suitably defined 
vacuum $|0>$~\cite{footnote}.
$ J^{\dagger}_{}$/$J^-_{}$ is the raising/lowering operator of the spin algebra.
and $\xi$ is a c-number. Alternatively, the coherent state can be 
represented as a point on the three-dimensional unit sphere, $\vec{n}$,
which is parameterized
by the two angles $\Theta$ and $\phi$;
$\xi=(1/2) \Theta e^{i\phi}$.

In Eq.~(\ref{eq:BFKM-action}),
the path integral runs over all periodic paths, i.e. 
$ \vec{n}(0)= \vec{n}(\beta)$.
$\omega(\vec{n})$ is a geometrical phase and equals the area on the unit sphere enclosed by
$ \vec{n}(\tau)$.
$s$ is the spin of the local moment.
In the following, we will be considering
the appropriate SU(N) or O(N)
generalizations 
of the SU(2) model.

{\it Model in the absence of the Berry phase:~~}
We consider first the quantum critical properties of the BFKM without 
the Berry phase term.
Simply removing the Berry phase term from the functional integral, Eq.~(\ref{eq:BFKM-action}),
results in an ill-defined measure.
Restraining however the spin path integral  to a subset of paths with
identical Berry phase, e.g. the subset of all great circles, 
the Berry phase term can be absorbed into the normalization of the partition function. 
After integrating out the bosonic bath, this leads to
\begin{eqnarray}
{\mathcal L}\,&=&\, i\mu(\tau)[\sum_{i}^{3}n_{i}^{2}-1] +J_K \vec{n}(\tau)
\cdot \bar{c}^{ }(\tau) \vec{\sigma} c^{}(\tau)\nonumber \\
 \,&+&\, \sum_{\sigma} \int_{0}^{\beta}
d\tau^{'} \bar{c}_{\sigma}^{ }(\tau)
G_{c}^{-1}(\tau-\tau^{'}) c_{\sigma}^{}(\tau^{'}) \nonumber
\\
\,&+&\,g^2 \int_{0}^{\beta}d\tau^{'} 
\vec{n}(\tau)G_{\phi}(\tau-\tau^{'})\vec{n}(\tau^{'}),
\label{L-no-berry-phase}
\end{eqnarray}
with $i\mu(\tau)$ being a
Lagrangian multiplier enforcing the constraint $\sum_{i}^{3}n_{i}^{2}=1$,
$J_K=sj_K/2$, and $g=s g_0$.
$G_c = - \langle T_{\tau} c_{\sigma\alpha}(\tau)
c_{\sigma\alpha}^{\dagger}(0) \rangle _0$,
and ${\cal G}_{\phi} =  \langle  T_{\tau} \phi(\tau)
\phi^{\dagger}(0) \rangle _0$.
The terms involving 
the electron 
fields 
are quadratic in them,
so the electron fields
can be exactly
integrated out as well:
\begin{eqnarray}
{\mathcal Z}\,&=&\, \int {\mathcal D}[\vec{n}]\, Det\Big[
G_c^{-1}
(\tau-\tau^{'})\delta_{\sigma \sigma^{'}}\! + \!J_K
\vec{n}(\tau)\cdot \vec{\sigma} \delta(\tau-\tau^{'})
\Big] \nonumber \\
&& \times \exp[-\int_{0}^{\beta} d\tau i\mu(\tau)[\sum_{i}^{3}n_{i}^{2}(\tau)-1]\nonumber \\
&& +g^2 \int_{0}^{\beta} d\tau \int_{0}^{\beta}d\tau^{'} 
\vec{n}(\tau)G_{\phi}(\tau-\tau^{'})\vec{n}(\tau^{'})].
\end{eqnarray}
With the help of $\exp[\ln Det M]=\exp[Tr \ln M]$
the effective action can be expressed as
\begin{eqnarray}
S\,&=&\, \int_{0}^{\beta} d\tau \, i\mu(\tau)[\sum_{i}^{3}n_{i}^{2}-1] \nonumber \\
&+&\, g^2  \int_{0}^{\beta} d\tau \int_{0}^{\beta}d\tau^{'} 
\vec{n}(\tau)G_{\phi}(\tau-\tau^{'})\vec{n}(\tau^{'})\nonumber \\
\,&+&\, 
Tr\ln(-G_c^{-1}\delta_{s,s^{'}}) \nonumber \\
\,&+&\,  
Tr\ln[{\bm 1} - J_K G_c \vec{n}(\tau)\cdot \vec{\sigma}_{s,s^{'}}
\delta(\tau-\tau^{'})].
\label{eq:eff-action}
\end{eqnarray}
The logarithm can be expanded in powers of 
$J_K G_c \vec{n}(\tau)\cdot \vec{\sigma}_{s,s^{'}}\delta(\tau-\tau^{'})$.
The odd powers in this expansion vanish,
since the Pauli matrices are traceless.

In order to systematically study
this action, we will first generalize
it
in such a way that the fluctuations around a saddle point vanish.
To do so, we extend the O(3) invariance 
of the action to 
an $O(N^2-1)$ symmetry,
with $\vec{n}$ 
containing $M \equiv N^2-1$ components. 
This corresponds to a generalization of Eq.~(\ref{L-no-berry-phase})
to the case with an 
$O(N^2-1)\times SU(N)$ symmetry.
We rescale the coupling constants $J_K$ and $g$ in terms of N,
so that a non-trivial large-N limit ensues.
The required rescaling of $J_K$ is determined by the N-dependence 
of the 
quadratic term
in the expansion of the logarithm of Eq.~(\ref{eq:eff-action}).
$Tr(J_K \vec{n}\cdot \vec{A} \,G^{}_{c})^2=
Tr(J_K  \sum_{i=1}^{N^2-1} n_i {\bm A}^{i} \,G^{ }_{c})^2$,
where the trace is over the (extended) spin space.
The ${\bm A}^{i}$ are Hermitean $N\times N$ matrices of unit determinant.
We  make use of the invariance of the trace and expand the generators in terms
of  N$\times$N matrices ${\bm B}^{l}$ ($l=1,\cdots,N^2$):  
  ${\bm A}^{i}=\sum_i^{N^2} a_{i,l} {\bm B}^{l}$. 
The set  ${\bm B}^{l}$ is chosen such,
that the $l$th matrix ${\bm B}^{l}$ has ${\bm B}^{l}|_{s,t}=e^{i\phi}$
and ${\bm B}^{l}|_{t,s}=e^{-i\phi}$ if and only if $l=N(s-1)+t$ with $s=1,\ldots,N$
and $t=1,\ldots,N$.
All other elements of ${\bm B}^{l}$ vanish identically.
The particular value of $\phi$ is left unspecified since it will only affect the
expansion coefficients $ a_{i,l}$.
$G_{c}$  is diagonal in the spin space.
Therefore, the  
quadratic term
should scale as $N\cdot M$. Rescaling $J_K^2 \rightarrow 
J_K^2/M=\tilde{J}_K^2$ 
renders the second term
proportional to $N$. Rescaling $g \rightarrow g/N=\tilde{g}$
has the same effect on the
similar term in the
the effective action involving the bosonic bath.
Finally, the constraint needs to be generalized 
to $\sum_i^{N} n_i^2=q_0N$.

Taking 
all these 
together, 
the large-N limit of the $O(N^2-1)\times SU(N)$ model
leads to a saddle-point equation
\begin{equation}
\chi^{-1}_{\mbox{\tiny loc}}
(\tau )
\,=\, \mu_0 
+ \tilde{J}_K^2 G^{}_{c,x}(\tau)G^{}_{c,\tilde{y}}(-\tau) 
+\tilde{g}^2 G_{\phi}(\tau) .
\label{ONsaddlepoint}
\end{equation}
At the saddle point, 
$i\mu(\tau)=\mu_0$
satisfies
\begin{equation}
\sum_{n}\chi^{}_{\mbox{\tiny loc}}(i \omega_n)=q_0 .
\end{equation} 
The second term of the RHS of Eq.~(\ref{ONsaddlepoint}) 
is just the particle hole bubble of 
the conduction electrons,
with 
$G^c_{x}(\tau)G^c_{\tilde{y}}(-\tau)\sim 1/\tau^2$. The long-time
behavior of $G_{\phi}(\tau)$ is 
specified 
by Eq.~(\ref{EQ:sub-Ohmic}), 
$ G_{\phi}(\tau)\sim 1/\tau^{2-\epsilon}$.
Solving the saddle point equation for a diverging 
$\chi_{\mbox{\tiny loc}}(\tau)$
results in a critical $\chi_{\mbox{\tiny loc}}$ with
$\chi_{\mbox{\tiny loc}}(\tau) \sim 1/\tau^{\epsilon}$,
implying
\begin{eqnarray}
\chi_{\mbox{\tiny loc}}(\omega,T=0) \sim 1/(-i\omega)^{1-\epsilon} .
\label{chi-omega-no-BP}
\end{eqnarray}

Away from the saddle point, i.e. for a finite $N$, additional 
interactions are present. The Lagrangian
multiplier $\mu$ acquires a $\tau$ dependence,
$i\mu(\tau)=\mu_0+\Delta\mu(\tau)/N$ to the sub-leading order.
This generates a new interaction vertex
of the form $\Delta\mu(\tau) n^2(\tau)/N$,
which gives rise to a quartic coupling 
of the field $n(\tau)$:
\begin{eqnarray}
\frac{u}{N}
\prod_{i}^{4} \int d\omega_i 
\delta(\sum_i^4 \omega_i)\prod_{i}^{4}n(\omega_i).
\end{eqnarray}
The scaling dimension of the field $[n(\omega)]=(2-\epsilon)/2$ 
follows from Eq.~(\ref{EQ:sub-Ohmic}).
As a result, the scaling 
dimension of the
quartic coupling
is
$[u]=2(\frac{1}{2}-\epsilon)$~\cite{Fisher.72}.
For $\epsilon<1/2$,
$u$ is a relevant perturbation and the low-energy
properties of the system will be governed by an interacting
fixed point with $u^*\neq 0$
and consequently hyperscaling and $\omega/T$-scaling.
This interacting fixed point is
the Ginzburg-Wilson-Fisher fixed point of the 
local 
$\phi^4$-theory.
For $1/2 \le \epsilon<1$, $u$ is irrelevant and 
will flow to zero; the Gaussian fixed point will be stable.
A vanishing quartic coupling makes the 
approach to the fixed point singular; in other words,
$u$ is dangerously irrelevant and therefore spoils
hyperscaling and $\omega/T$-scaling.
In the context of the long-ranged Ising model this process has been
discussed 
recently in~\cite{Kirchner.08b}.
The dangerously irrelevant coupling leads to
\begin{eqnarray}
\chi_{\mbox{\tiny loc}}(\omega=0,T) \sim 1/T^{\frac{1}{2}}
\label{chi-T-no-BP}
\end{eqnarray}
Comparing Eqs.~(\ref{chi-omega-no-BP},\ref{chi-T-no-BP}) shows 
that $\chi_{\mbox{\tiny loc}}(\omega,T)$ disobeys
an $\omega/T$ scaling.

{\it Model in the presence of the Berry phase:~~}
\begin{figure}[t!]
\includegraphics[width=0.5\textwidth]{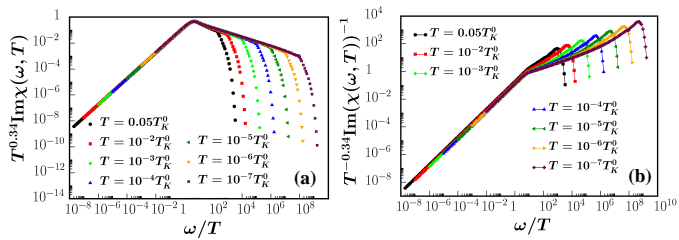}
\caption{Scaling plot of (a) $T^{0.34}\mbox{Im}
\chi_{\mbox{\tiny loc}}(\omega,T)$ and 
(b) $T^{-0.34}\mbox{Im}\big(1/\chi_{\mbox{\tiny loc}}
(\omega,T)\big)$ for $\epsilon=2/3$.
Numerical parameters
are specified in the main text.
}
\label{FIG1}
\end{figure}
In order to generalize SU(2) to SU(N) within a path integral formulation it is necessary to
use proper coherent states over SU(N). Such coherent states can be constructed in analogy
to the SU(2) case and a corresponding Berry phase term for the SU(N) spin path integral with 
similar topological properties emerges~\cite{Nemoto.00,Mathur.02,Read.89}. 
This model in the presence of the Berry phase term
can be studied in a dynamical large-$N$ limit~\cite{Zhu.04}
of the SU(N)$\times$ SU($\kappa$ N) BFKM:
\begin{eqnarray}
{\cal H}_{\text{MBFK}} &=&
({J_K}/{N})
\sum_{\alpha}{\bf S}
\cdot {\bf s}_{\alpha}
+ \sum_{p,\alpha,\sigma} E_{p}~c_{p \alpha
\sigma}^{\dagger} c_{p \alpha \sigma}
\nonumber\\
&+&
({g}/{\sqrt{N}})
{\bf S} \cdot
{\bf \Phi}
+ \sum_{p}
w_{p}\,{\bf \Phi}_{p}^{\;\dagger}\cdot {\bf \Phi}_{p},
\label{H-MBFK}
\end{eqnarray}
where  $\sigma = 1, \ldots, N$ and
$\alpha=1, \ldots, M$ 
are the spin and channel indices respectively,
and ${\bf \Phi} \equiv \sum_p ( {\bf \Phi}_{p} +
{\bf \Phi}_{-p}^{\;\dagger} )$ contains $N^2-1$
components.  
The local moment is  expressed in terms of pseudo-fermions
$S_{\sigma,\sigma^{\prime}}=f^{\dagger}_{\sigma}f^{}_{\sigma^{\prime}}
-\delta_{\sigma,\sigma^{\prime}}Q/N$,
where $Q$ is related to the chosen irreducible representation 
of SU(N)~\cite{Parcollet.98,Cox.93}.
The quartic term between conduction electrons and pseudo-fermions is expressed in terms of
a bosonic
decoupling field $B_{\alpha}$.
The large-$N$ saddle-point equations are
\begin{eqnarray}
\Sigma_B(\tau) &=& - G_{c}(\tau) G_f(-\tau);
 \nonumber \\
\Sigma_f(\tau)&=& \kappa G_{c}(\tau) G_B(\tau) + g^2
G_f(\tau){\cal G}_{\Phi}(\tau); \nonumber \\
G_B^{-1}( i\omega_n) &=&
1/{J_K} - \Sigma_B( i\omega_n); \nonumber \\
G_f^{-1}(i\omega_n)& =& i\omega_n - \lambda -
\Sigma_f(i\omega_n);
\label{NCA}
\end{eqnarray}
together with a constraint $G_f(\tau\rightarrow 0^{-})=Q/N$.
Here, $\kappa=M/N$ and $\lambda$ is a Lagrangian multiplier.
The analytically continued equations  (\ref{NCA}) can
be self-consistently solved for any
frequency ($\omega$) and temperature ($T$)~\cite{Zhu.04}.
Eqs.~(\ref{NCA}) completely capture
the (full) quantum dynamics of the problem and
contain the effect of
the Berry phase in the path integral approach.
At the QCP, the order parameter susceptibility~\cite{Zhu.04}
behaves as
\begin{equation}
\chi_{\mbox{\tiny loc}}(\omega,T=0)\sim 1/\omega^{1-\epsilon};~~ 
\chi_{\mbox{\tiny loc}}(\omega=0,T)\sim 1/T^{1-\epsilon}
\end{equation}
for all $0<\epsilon<1$.
Fig.~\ref{FIG1}(a) 
shows the the $\omega-T$-scaling of 
$\chi_{\mbox{\tiny loc}}(\omega,T)$
for $\epsilon=2/3>1/2$.
The numerical parameters are
$\kappa =1/2$,
$Q/N=1/2$, and
 $N_0(\omega) = (1/\pi){\rm exp}(-\omega^2/\pi)$
for the conduction electron density of states.
The nominal bare Kondo scale is
$T_K^0 N_0(0)\equiv {\rm exp}(-1/N_0(0)J_K) \approx 0.06$,
for fixed $J_K N_0(0)=0.8$.
The bosonic bath spectral function
$\sum_{p} \delta(\omega-w_{p}) \sim \omega^{1-\epsilon}_{}$
is cut off smoothly
at $2 \omega_c N_0(0)\approx 0.05$.

\begin{figure}[t!]
\includegraphics[width=0.5\textwidth]{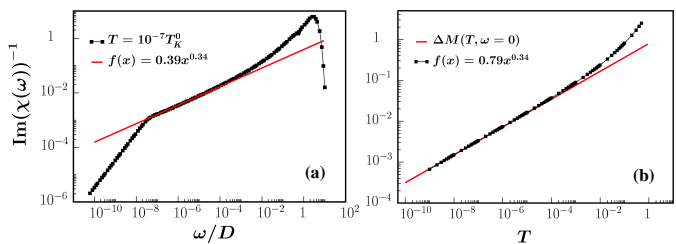}
\caption{(a) $\mbox{Im}\big(\chi^{-1}_{\mbox{\tiny loc}}(\omega,T)\big)$
 displays power law behavior $\sim \omega^{1-\epsilon}$ for $\omega>T$. 
(b) temperature dependent part of the 
saddle point self-energy $\Delta M(T,\omega=0)\sim T^{1-\epsilon}$.}
\label{FIG2}
\end{figure}

The  $\omega/T$-scaling of $\chi_{\mbox{\tiny loc}}(\omega,T)$
for $1/2 \le \epsilon < 1$ implies the breakdown of the quantum
to classical mapping, since the mapped classical critical point
does not obey $\omega/T$-scaling for this range of $\epsilon$.
An important issue is whether the
SU(N)$\times$ SU($\kappa$ N) result anchored around large 
N can be extended to a finite N or whether it is susceptible
to the presence of dangerously irrelevant couplings.
In order to address this question we introduce a
self-energy for the order parameter
susceptibility,
\begin{equation}
M(\omega,T)\equiv g_c^2\chi_0^{-1}(\omega) 
-1/\chi_{\mbox{\tiny loc}}(\omega,T),
\end{equation}
where $\chi_0^{-1}(\omega)$ follows from Eq.~(\ref{EQ:sub-Ohmic}) 
and $g^2_c(J_K^{})$ is the value of the
coupling constant $g^2$ at which the system becomes 
critical for a given $J_K^{}$. 
The self-energy $M(\omega,T)$ thus
defined will be $T$-dependent, 
since  $\chi_0^{-1}(\omega)$ is $T$-independent but the
inverse of $\chi_{\mbox{\tiny loc}}(\omega,T)$ shows 
non-trivial $\omega/T$-scaling, see Fig.~\ref{FIG1}(b).
[By contrast, at the Gaussian fixed point arising in the case
without the Berry phase term, described in the previous 
section, the corresponding self-energy
$M(\omega,T)$ at $N=\infty$ is $T$-independent.]
In Fig.~\ref{FIG2}(a) we show 
$\mbox{Im}(1/\chi_{\mbox{\tiny loc}}(\omega,T_1))$ for 
$T_1=10^{-7}T_K^0$. It shows power-law behavior over 
several decades of frequency, with the
exponent very close to $1-\epsilon=1/3$. 
The
power-law behavior is cut off at around $\omega \sim T_1$.
The constant part of the self-energy $M(\omega=0,T=0)$
determines the critical value of $g^2$.
The temperature and frequency dependent
part of the self-energy, 
$\Delta M(\omega,T)=M(\omega,T)-M(\omega=0,T=0)$,
is shown in Fig.~\ref{FIG2}(b) for $\omega=0$. 
It has the important property
\begin{equation}
\Delta M(\omega=0,T)\sim T^{1-\epsilon}.
\end{equation}
Since the exponent in this temperature dependence is
the same as that of its frequency dependence at $T=0$,
it cannot be modified by any subleading temperature-dependent
terms that could possibly be generated by a dangerously
irrelevant coupling at a finite N.
[By contrast, in the absence of the Berry phase term,
the temperature dependence $\Delta M(\omega=0,T)$
is entirely determined by the $T^{1/2}$ term associated with
the $1/N$ corrections.]
This implies that the observed $\omega/T$-scaling,
even for $1/2 \le \epsilon<1$,
survives beyond 
the large-$N$ saddle point of the SU(N)$\times$ SU($\kappa$ N) BFKM,
extending to subleading orders in $1/N$.

The qualitative difference caused by the Berry phase term
implies that the quantum interference effect is important
for the universal properties of the QCP in the 
spin-isotropic BFKM.
This is a natural manifestation of the fact that the Kondo effect
-- the formation of a Kondo singlet state and its critical destruction
-- involves the quantum entanglement of the local moment and spins
of the conduction electrons.
By bringing out the explicit effect of the Berry phase term,
the spin-isotropic BFKM represents a prototype case in which
the breakdown of the quantum-to-classical mapping can be studied.
The situation in the Ising-anisotropic 
cases~\cite{Vojta.05,Glossop.05,Kirchner.08b,Winter.08}
is not as clear-cut. Still, since the SU(2) symmetry is
restored in the Kondo screened phase of any spin-anisotropic
Kondo system, it is plausible that related effects also come
into play in the quantum criticality of the spin-anisotropic 
dissipative Kondo-like models.

In summary, we have analyzed the influence of the Berry phase,
the topological phase term of the
spin path integral, on the quantum critical properties of the 
Bose-Fermi Kondo model. We have done so using 
large-$N$ approaches based on appropriate generalizations
of the model with and without the Berry phase term,
and taking into account effects beyond the leading 
order in $1/N$.
Without the Berry phase term, an interacting fixed point 
with
the $\omega/T$-scaling occurs for $0<\epsilon<1/2$
but a Gaussian fixed point spoiling the 
$\omega/T$-scaling arises for $1/2 \le \epsilon<1$.
With the Berry phase term, the  Bose-Fermi
Kondo model shows an interacting fixed point with
$\omega/T$-scaling over the entire range $0<\epsilon<1$.

We thank C.~J.~Bolech, A.~Inomata, S.~Yamamoto and L.~Zhu  for useful discussions.
This work has been supported in part by
NSF Grant No. DMR-0706625, the Robert A. Welch Foundation, 
the W. M. Keck Foundation,
and the Rice Computational Research Cluster
funded by NSF
and a partnership between Rice University, AMD and Cray.



\begin{thebibliography}{19}
\bibitem[{\citenamefont{footnote}()}]{footnote}
\bibinfo{note}{\bibfnamefont{$|0>$ for SU(2) coherent states can be any of the 2s+1 states  $|s,m>$ with 
$m=-s,\ldots,s$. Customary choices are $|0>=|s,s>$ or $|0>=|s,-s>$, so that the coherent states
can be generated by either $J^-$ or $J^\dagger$ alone. The determination of the
normalizing factor 
which gives rise to the 
Berry phase in the path integral requires the proper use of the  Baker-Cambell-Hausdorf formula
in analogy to bosonic coherent states, 
$T(z)=e^{[za^\dagger - \bar{z} a]}=e^{-|z|^2/2}e^{za^\dagger}e^{\bar{z} a}$. }}
\end{thebibliography}
\end{document}